\documentclass{svproc}

\usepackage{xcolor} 
\usepackage{url}

\usepackage[utf8]{inputenc}
\usepackage{amsmath}
\usepackage{amsfonts}
\usepackage{amssymb}
\usepackage{verbatim}

\usepackage{float}
\usepackage{xcolor}
\usepackage{framed}
\usepackage{rotating}
\usepackage{afterpage}
\usepackage{url}

\usepackage{cite}

\colorlet{shadecolor}{gray!15}

\newcommand{\cling}{P}    
\newcommand{\Units}{\mathbf{S}}
\newcommand{\so}{s}
\newcommand{\So}{S}
\newcommand{\Sv}{Y}

\newcommand{\Ldrs} {{\cal R}}

\newcommand{\x}{\mathbf{x}}

\newcommand{\leadl}{\mathbf{r}}

\newcommand{\spec}{\mathrm{spec}}

\def\argmin{\mathop{\rm arg\ min}\nolimits}

\newcommand{\Cf}{\text{Err}}
\newcommand{\err}{\text{err}}

\newcommand{\R}{R}
\newcommand{\rr}{r}
\newcommand{\w}{w}

\newcommand{\y}{\mathbf{y}}

\floatstyle{ruled}
\floatname{algorithm}{\textbf{ALGORITHM}}

\begin{document}
\mainmatter 
%
\title{Identifying patterns of main causes of death \\ in the young EU population}
\titlerunning{Mortality cause patterns in young people in the EU}  
%
\author{Simona Korenjak-\v{C}erne\inst{1} \and Nata\v{s}a Kej\v{z}ar\inst{2}}
\authorrunning{Korenjak-\v{C}erne et al.} 
%
\tocauthor{Simona Korenjak-\v{C}erne and Nata\v{s}a Kej\v{z}ar}
\institute{University of Ljubljana, School of Economics and Business, \\ and Institute of Mathematics, Physics and Mechanics, Slovenia\\
\email{simona.cerne@ef.uni-lj.si}, 
\and
University of Ljubljana, Faculty of Medicine, \\ Institute for Biostatistics and Medical Informatics, Slovenia \\
\email{natasa.kejzar@mf.uni-lj.si}}

\maketitle              

\begin{abstract}
The study of mortality patterns is a popular research topic in many areas. We are particularly interested in mortality patterns among main causes of death associated with age-gender combinations. We use symbolic data analysis (SDA) and include three dimensions: age, gender, and patterns across main causes of death. In this study, we present an alternative method to identify clusters of EU countries with similar mortality patterns in the young population, while considering comprehensive information on the distribution of deaths among the main causes of death by different age-gender groups. We explore possible relationships between mortality patterns in the identified clusters and some other sociodemographic indicators.
We use EU data of crude mortality rates from 2016, as the most recent complete data available.

\end{abstract}

\section{Introduction}

Mortality data on causes of death provide important information about the public health in the observed geographic area. Consequently, the study of mortality patterns is a very popular research topic in many areas (e.g., demography, health economics, public policy, and actuarial science), and therefore several approaches to the analysis of these data have been developed depending on specific needs. For an overview of these approaches with many references, see for example the paper of Van Raalte \cite{surveyPopStud-21}. As highlighted in this paper, most life course studies are still conducted at the individual level, while the main challenge is to link life course studies more closely to mortality patterns at the aggregate level \cite{surveyPopStud-21} (p. S107). In this paper, we present a possible approach to the analysis of mortality data at the aggregate level.

Our exploratory study focuses on the leading causes of death in the young population. Our study was motivated by the Centres for Disease Control and Prevention (CDC) publication summarising 10 leading causes of death by age group in the United States. The three leading causes of death among the young population in the United States are unintentional injuries, suicide, and homicide. In the fourth place are malignant neoplasms. Since the most common causes of death in the young population aged 20-39 are external (i.e., accidents, especially traffic accidents, suicides, and assaults), which are preventable, we focus on mortality patterns in the young population in EU countries.

It is well-known that the number of deaths is strongly related to age, gender, and cause of death. 
All three factors can be captured by selecting all possible combinations of these factors as variables and then applying classical multivariate data analysis techniques. However, in this case internal relationships among these variables are usually not taken into account. These relationships 
may be described with an empirical distribution of deaths among different causes of death. To include them simultaneously in the analysis, more advanced analysis methods are needed.

The data that we deal with are aggregated deaths, and we analyze them using symbolic data analysis (SDA) \cite{bo-did-00, bi-did-06, did-nor-08}. SDA methods consider internal variability; in our case, variability among deaths by cause of death. Following this path, we use symbolic data descriptions  \cite{EBR-19} and consider data as symbolic objects, where the units of interest are EU countries and the symbolic variables are five-year age-gender groups described with distributions of the (\textit{expected}) number of deaths
across selected causes of death. We name these distributions \textit{mortality patterns across causes of death}. To find groups of EU countries with similar mortality patterns, we adapt the hierarchical and clustering method presented in \cite{NKSKCVB-21} and implemented in the R package {\tt clamix}, which we modify to include the requirements of our specific problem.

In addition to preserving the relationship structure, another question that we want to tackle is how to sufficiently transform mortality rates to make them directly comparable between countries.

The objectives of this study are twofold: (1) to present an alternative method to identify clusters of EU countries with similar mortality patterns in the young population that takes into account more comprehensive information on the three dimensions of the data: age, gender, and distribution across causes of death; and (2) to explore possible relationships between mortality patterns in the identified clusters and some other sociodemographic indicators, which can serve as a starting point for further detailed quantitative and qualitative investigations of possible associations. 
We use EU data on crude mortality rates from 2016, which are the most recent complete data available.

The rest of the paper is organized as follows. In the second section, we present an additional rationale for our data selection and its preparation. In the third section, we explain the adaptations of the clustering method to our symbolic descriptions of the data. In the fourth section, we present and comment on some of the results. The final section will give the concluding remarks of this study.

\section{Data}

The young population is generally considered to be very healthy, and therefore the leading causes of death in this population are often associated with risky behaviors. The leading causes of death in the general population in EU countries in 2016 and 2017 were circulatory diseases and various cancers; followed by respiratory diseases; while external causes of death, including accidents, suicides, homicides, and other violent causes of death, ranked fourth \cite{eurostat-20a, oecd-20}. 
The leading causes of death in the young population are very different \cite{eurostat-20b}, with external causes being by far the most common causes of death (the percentage varies by age group but usually comprises more than half of the deaths).  

The boundaries of the age groups that comprise the young population are not clearly defined. 
In this work, we focus on the ages from 20 to 39 years because this is the time when most people face major changes in their lives (usually starting a new job, becoming independent from parents, finding a new place to live, planning own family, raising children, etc.). This selection also coincides with the psychological developmental stage of young adulthood that was introduced by Erikson \cite{erikson-94}.
An additional reason for this particular data selection is that the generation of ages from 20 to 39 years is often overlooked in specific health studies because their causes of death are associated with risky behaviors that are often considered to be personal choices. Meanwhile, risky behaviors are closely related to mental status, with stress being one of the most important factors. Most of these causes of death are preventable, and consequently we should not overlook these important issues in our society. It is therefore important to develop appropriate prevention programs for this target group because they face very specific challenges in life and represent about a quarter of the total population of the EU. 

\subsection{Symbolic representation of mortality data}

For easier understanding and interpretation, mortality data are usually presented with {\it crude death rate} (CDR), which is calculated as the ratio between the (total) number of deaths and the (total) population size, and is usually expressed per 100,000. Thus, it represents the number of deaths per 100,000 persons at risk in the observed region. Because mortality data are affected by age and gender, age- and gender-adjusted death rates are more appropriate for comparisons between regions with different population age and gender structures. Another factor strongly associated with age and gender, in addition to the number of deaths, is the cause of death.

Given the key recommendations in the work of Anderson and Rosental \cite{An-Ros-98} to obtain as much of the age- and gender-cause-specific information and to make the data comparable across countries, we decided to convert the observed number of deaths to the \textit{expected} number of deaths. To do this, we consider age- and gender-specific death rates for a common (standard) population, which we also recalculated on a two-dimensional (age-gender) structure for this purpose.  

We took the latest complete mortality data that was available from Eurostat, which is the statistical office of the EU  \cite{eurostat-data}, for the year 2016. Thus, we originally dealt with 28 units (i.e., EU countries) and eight symbolic variables (for each gender we had four five-year age groups: 20-24, 25-29, 30-34, and 35-39). 

We considered the data as symbolic objects \cite{bo-did-00, bi-did-06, did-nor-08}, where the units of interest are the EU countries and the symbolic variables are five-year age-gender groups related to young adults, which are described with the distributions of deaths among the selected causes of death. They are called "symbolic" because they represent aggregations of deaths and are described with adjusted "symbolic" variables that contain information on the number of deaths and their distributions among causes of death. Some of this information would be lost if the data were presented in a classical way.

\subsubsection{Data description with distributions}

We consider each country as an observation unit (class) $\so_i$ and describe it with a symbolic description $S_i$. Each country is in our case described with eight symbolic variables $Y_j, j=1, ..., 8$, representing mortality data for the following age-gender groups: 
$Y_1$: 20-24 male, $Y_2$: 20-24 female, $Y_3$: 25-29 male, $Y_4$: 25-29 female, $Y_5$: 30-34 male, $Y_6$: 30-34 female, $Y_7$: 35-39 male, and $Y_8$:  35-39 female. 

More generally: 
a unit  $S_i$ 
from a data set of $n$ units $\Units$ is a list of values of modal valued variables $Y_j, \; j = 1,\ldots,p$. A symbolic variable $Y_j$ has $m_j$ categories (subsets). Each variable is described by a list of values of corresponding categories:
\begin{equation*}
	\Sv_j = \left [  L_{j1}:y_{j1}, \ldots, L_{jm_j}:y_{jm_j} \right ].
\end{equation*}
In the description, $L_{j\ell}$ is a category label and $y_{j\ell}$ is its value. 
 
In our case, values $L_{j\ell}$ are the causes of death and $y_{j\ell}$ are the proportions of deaths by the corresponding cause. We have the same categories for all age-gender combinations ($m_j = 7$ for all eight symbolic variables $j, j=1,...,8$). 
We selected leading causes of death for EU countries in the age groups 20 to 39 years. These are 
\begin{itemize}
   \item Neoplasms (\emph{Neop}), 
   \item Nerves (\emph{Nerv}) (Diseases of the nervous system and sense organs (G00-H95)), 
   \item Circulatory (\emph{Circ}) (Diseases of the circulatory system (I00-I99)), 
   \item Respiratory (\emph{Resp}) (Diseases of the respiratory system (J00-J99)), 
   \item Accidents (\emph{Acc}) (V01-X59, Y85, Y86) and 
   \item Suicides (\emph{Suic}) (Intentional self-harm). 
\end{itemize}
The codes in parentheses refer to the ICD classification \cite{ICD-10}. We did not include homicide as a separate cause because it represents a very small part of the causes of death among the young population in EU countries. \bigskip 
 
A description of symbolic object $\so_i$ is a vector of modal values $\So_i = [ \y_{i1}, \ldots, \y_{ip} ]$, where
$\y_{ij} =  \left [ L_{j1}:y_{ij1}, \ldots, L_{jm_j}:y_{ijm_j} \right ]$
is the modal value corresponding to symbolic variable $Y_j$ for symbolic object $\so_i$. 
\medskip

For example, if $\so_i$ = EU28, the whole European union, and $Y_j$ = $Y\textsubscript{20-24 male}$, then 

$\small \y\textsubscript{EU28, 20-24 male} = $\\
$\tiny = \left [ \text{Neop}:0.093, \text{Nerv}:0.046, \text{Circ}:0.054, \text{Resp}: 0.022,\text{Acc}: 0.387, \text{Suic}: 0.218, \text{Oth}:0.180 \right ]. $ \\

From this description we clearly see that the two leading causes of death in the male population in the age group from 20 to 24 years are accidents ($38.7 \%$) and suicides ($21.8 \%$).

Given that for a symbolic variable $Y_j$ the categories are fixed, we fix their order and omit them from the modal value list---the category is determined by the value's position
$\y_{ij} =  \left [ y_{ij1}, \ldots, y_{ijm_j} \right ]$.
For the example above, we have: \\

$\small \y\textsubscript{EU28, 20-24 male} = [0.093,0.046,0.054,0.022,0.387,0.218,0.180]$. \\

Note that in our case we considered the following constraint $\sum_{\ell=1}^{m_j} y_{ij\ell} =1$ for all symbolic objects $\so_i, i=1,..., n$, and all symbolic variables $Y_j, j=1,...,p$. If we do not include any additional information (e.g., weights representing levels, such as the number of deaths), then our symbolic variables for each age-gender combination represent compositions \cite{pawlowsky2011compositional}. Each of these compositions represents the \textit{mortality pattern} across causes of death, which are related to the specific age-gender combination. \label{mpattern}

\subsubsection{Selection of weights}

Age-gender-specific rates were obtained from the Causes of Death -- Crude Death Rate table by NUTS 2 regions of residence, 3-year average \cite{eurostat-data}. 
We intended to compare mortality rates across EU countries. However, each EU country has its own age-gender structure. Thus, we had to convert these rates. We used a standard population for this purpose.

Unfortunately, the ESP (European Standard Population) that is used by Eurostat is only one-dimensional; that is, only for age groups, a breakdown by gender is unavailable \cite{RESPR-13} (p. 15).
Therefore, we used the standard population from the Federal Health Monitoring Information System website \cite{ISFHM} as a basis and recalculated the separate male and female age distribution of the new European Standard Population into a two-dimensional standard population by age and gender.

The calculation follows the basic idea of the age-adjusted death rate, which is based on an age-gender-specific death rate (i.e., the number of deaths per 100,000 people in the age-gender group relative to the standard population). 
We define weight $w_{xj}$ as the \textit{expected} number of deaths in the age-gender group $j$ in country $X$ by
\begin{equation} \label{weights}
   w_{xj} = \frac{D_{xj}}{P_{xj}} \cdot P_{std,j}, 
\end{equation}   
where $P_{std,j}$ denotes the number of persons in the age-gender group $j$ in the standard population, $D_{xj}$ is the number of deaths in the age-gender group $j$ in country $X$, and $P_{xj}$ the number of persons in the age-gender group $j$ in the country $X$.

A description of mortality data for the entire young population in the EU is given in Appendix 1. We easily see the differences in weight by age and gender, as well as different mortality patterns in terms of causes of death in different age-gender combinations.

\section{Adapted clustering methods}

We used the clustering method from \cite{NKSKCVB-21} to obtain clusters of EU countries with similar patterns of the main causes of death in the young population based on the presented data description, which we  adjusted with our data constraints. For this purpose, we adapted the R package {\tt clamix} to our specific problem.

In the description of a general model for the leader method  \cite{NKSKCVB-21}, where the set of feasible clusterings is 
\emph{a set of partitions} of the finite set of units $\Units$ in $k$ clusters, the criterion function is the sum of all cluster "errors". The "error" of a cluster $\err(C)$ is the sum of the dissimilarities of its units with respect to the optimal representative, which is called the leader $\R_C$ of the cluster:
\begin{equation*} \label{CritF}
   \Cf (\cling) = \sum_{C \in \cling} \err(C), \textrm{where } \cling \textrm{ is a partition and }
   \err(C) = \sum_{X \in C} d(X,\R_C).
\end{equation*}

Let the description of the unit $X$ (representing the symbolic object $\so$) for the symbolic variable $Y_j, j=1,...,p,$ be a pair $(w_{xj},\x_{j})$, where \break
$\x_{j} = \left [ L_{j1}:x_{j1}, \ldots, L_{jm_j}:x_{jm_j} \right ]$, $\sum_{\ell=1}^{m_j} x_{j\ell} = 1$. If the order of categorical values is fixed, then it can be simplified to 
$\x_{j} = \left [ x_{j1}, \ldots, x_{jm_j} \right ]$. The weights $w_{xj}$ are related to the unit $X$ and the $j$-th variable. 

Compared to the more general formulation in \cite{NKSKCVB-21}, here we have an additional condition for each symbolic variable; that is, the sum of the values equals 1.
	
For each variable $Y_j, j=1, ..., p,$ we assume that the leader $\R$ is represented by the list of non-negative vectors of size $m_j$:
$ 
	\leadl_j = \left [ \rr_{j1}, \ldots, \rr_{jm_j}\right ], \mbox{ where } 0\leq \rr_{j\ell} \leq 1 \mbox{ for all } \ell, \ell = 1, ..., m_j, \mbox{ and  (again)} \sum_{\ell=1}^{m_j} \rr_{j\ell} = 1.
$
We denote the set of all possible representatives by $\Ldrs$.
The representation space is due to the constraints limited to the $\Ldrs = [0,1]^{m_1} \times [0,1]^{m_2} \times \dots \times [0,1]^{m_p}$. 

We choose the squared Euclidean distance for the basic dissimilarity 
because it allows for easy computation of the leaders while emphasizing the largest value
and also because there are zeros in some categories in some EU countries that we wanted to retain.
Thus, in our case, the dissimilarity measure between the unit $X$ and the cluster representative $\R$ is defined as 
\begin{equation*}
 d(X,\R) =   \sum_{j=1}^{p}  w_{xj}  d_E^2(\x_{j},\leadl_j) = \sum_{j=1}^{p}  w_{xj} \sum_{\ell=1}^{m_j} ||x_{j\ell} - \rr_{j\ell}||^2, \quad \w_{xj} \geq 0, 
\end{equation*}
where $d_E^2(\x_{j},\leadl_j)$ is the squared Euclidean distance 
and $\w_{xj}$ are weights with respect to the description of the unit $X$ by the $j$-th symbolic variable.

The new leader $\R_C$ of the cluster $C$ is determined with the optimization with constraints
\[ \R_C = \argmin_\R  \sum_{X \in C} d(X,\R) =
\big[ \argmin_{\leadl_j} \sum_{X \in C} w_{xj}  d_E^2(\x_{j},\leadl_j)  \big]_{j=1}^p.
\]

The optimal representative, which is the leader of the cluster $C$, is described by the symbolic variable $Y_j$, which is described by the vector $\leadl_j =[ \rr_{j1},\ldots,\rr_{jm_j} ]$, \\
where 
\begin{equation*}
\rr_{j\ell} = \frac{1}{\sum_{X \in C} w_{xj}} \sum_{X \in C} w_{xj} x_{j\ell} \mbox{\;\;\;for all \;\;} \ell, \ell=1, ..., m_j.
\end{equation*}
Thus, the components are the weighted average of $\ell$-th components $x_{j\ell}$ over all units $X$ in cluster $C$.  
Its derivation can be found in Appendix 2.

The dissimilarity in the agglomerative hierarchical clustering compatible with the criterion function $\Cf (\cling) = \sum_{C \in \cling} \err (C)$ is \cite{NKSKCVB-21}:  
\begin{equation*}
 D(C_u,C_v) = \err (C_u \cup C_v) - \err (C_u) - \err (C_v),
 \end{equation*}
and the dissimilarity between the merged clusters for our chosen data description and the previously defined optimization problem with the optimal cluster representative is
\begin{equation*}
D(C_u,C_v) = \frac{1}{p} \sum_{j=1}^{p} \frac{w_{uj}\cdot w_{vj}}{w_{uj} + w_{vj}} d_E^2(\mathbf{u},\mathbf{v}), 
\end{equation*}
where $\displaystyle \mathbf{u} \mbox{ and } \mathbf{v}$ are the leaders of the clusters $C_u$ and $C_v$, respectively, \\
and $\displaystyle w_{uj} = \sum_{X \in C_u} w_{xj}, w_{vj} = \sum_{X \in C_v} w_{xj}$. If we are only interested in mortality patterns, then we select  $w_{xj} = 1$ for all countries $X$ and all symbolic variables $V_j, j= 1, j=1,...,p.$ 

\section{Results}

In this section, we present some of the results obtained with the clustering methods adapted to our symbolic data descriptions. 
In particular, we present the results obtained with weights (see Eq.~\eqref{weights}).
The data consist of $n = 28$ EU countries (classes) in 2016. To reduce random variation, we considered mortality rates as 3-year averages for the number of deaths.
We have $p = 8$ symbolic variables corresponding to eight age-gender combinations, based on four five-year age groups for young adults: 20-24, 25-29, 30-34, and 35-39; $m_j = m = 7$.  
The selected categories of the six leading causes of death for EU countries for young adults (four diseases and two external causes, listed with ICD-10 codes) are: 
	\begin{itemize}
		{\small 
			\item[] DISEASES: \\
		 Neop -- Neoplasms (C00-D48), \\
		 Nerv --  Diseases of the nervous system and sense organs (G00-H95), \\
		Circ -- Diseases of the circulatory system (I00-I99), \\
		Resp -- Diseases of the respiratory system (J00-J99), 
		   \item[] EXTERNAL CAUSES OF DEATH: \\
		Acc -- Accidents (V01-X59, Y85, Y86), and \\
		Suic -- Suicides (Intentional self-harm (X60-X84, Y87)), 
		   \item[] OTHER CAUSES \\
		Oth -- all other causes of death fall into a final category called Oth.
	}
	\end{itemize}
		
\begin{figure}
\vspace*{-1 cm}
  \includegraphics[width=12.5cm, height=6cm]{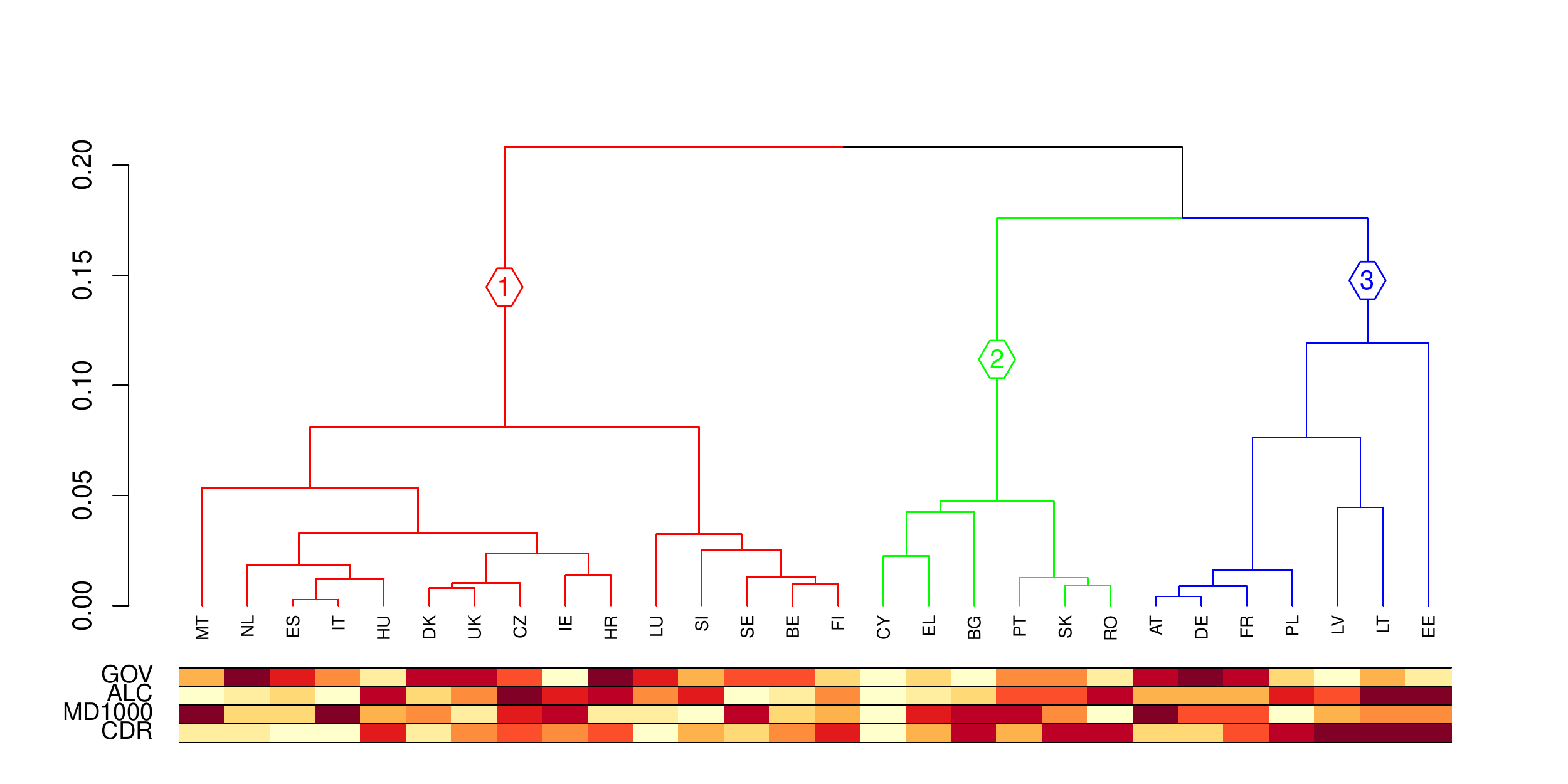}
    \caption{Dendrogram of 28 EU countries based on the differences in mortality patterns across causes of death with some sociodemographic factors displayed at the bottom}
   \label{dendro}
\end{figure}

From the dendrogram shown in Fig.~\ref{dendro}, we can easily identify three large (although not very homogeneous) clusters. 
From GapMinder \cite{GapMinder}, we obtained and plot under the dendrogram the following external sociodemographic factors that could be related to mortality:
\begin{itemize}
		{\small 
			\item[] MD1000 -- Medical doctors per 1000 people (last year in the data 2006), 
			\item[] ALC -- Alcohol consumption per adult (last year in the data 2008), and 
			\item[] GOV -- Government health spending of total gov spending percent (last year in the data 2010). \medskip
			\item[] We added the number of \textit{expected} deaths in the entire age group 20-39 years per 100,000 of the standard population (CDR).
	}
	\end{itemize}	

The values are replaced by colours (empirical deciles), with a darker colour representing a higher value. The clustering procedure did not include sociodemographic factors.

\begin{figure}[h!]
   \includegraphics[width=12cm, height=7cm]{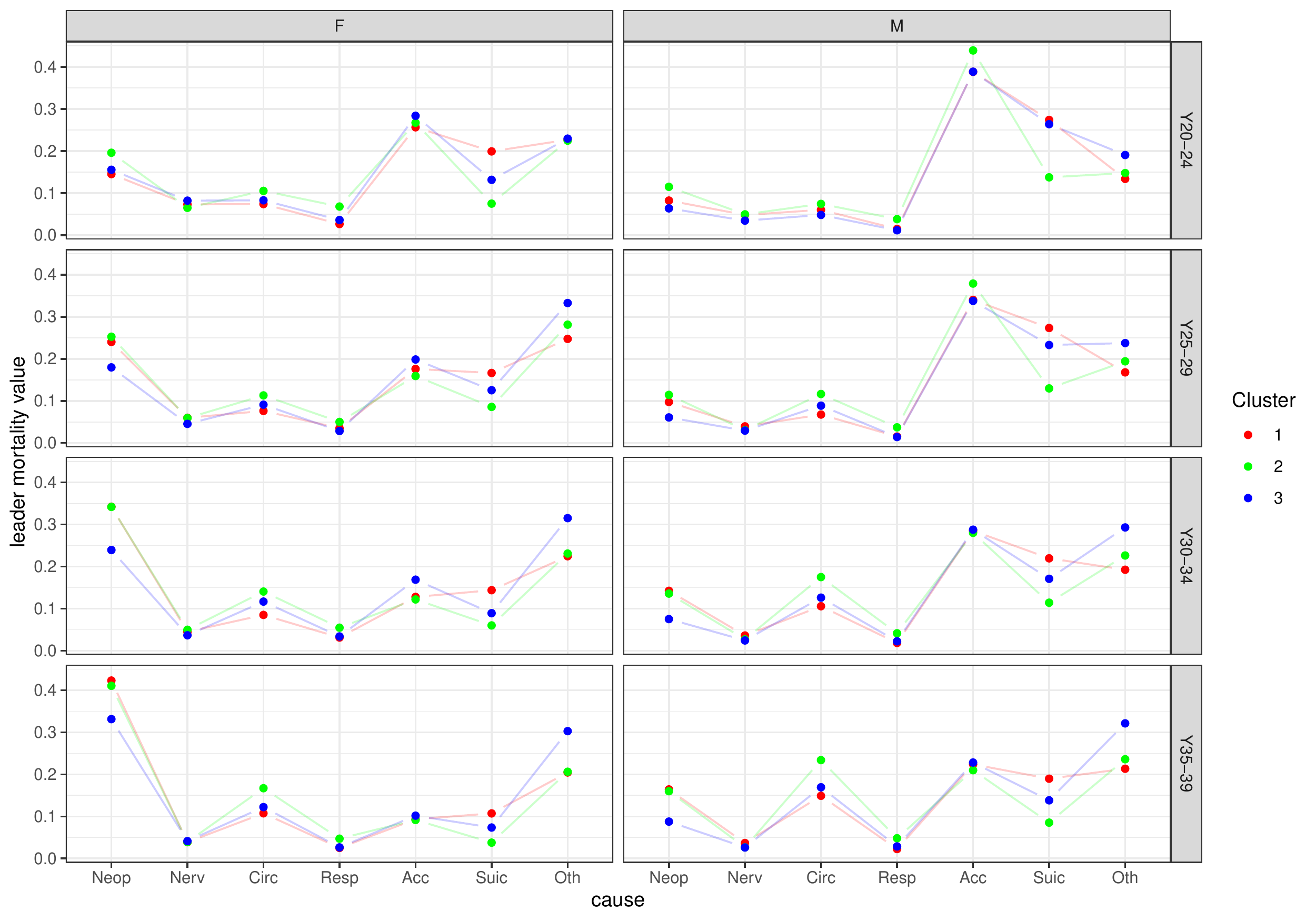}
   \caption{\textit{Mortality patterns across causes of death} for eight age-gender combinations for three main clusters. Note: The order of the categories is fixed. The graphs should not be regarded as classic line charts -- broken lines are only used to facilitate comparison of \textit{mortality patterns}.}
    \label{leaders}
\end{figure}

The \textit{mortality patterns across causes of death} of cluster leaders for three main clusters are shown in Fig.~\ref{leaders}, where we see the largest absolute differences in neoplasms, circulatory diseases, suicides, and the other category. 

Given that the cluster leader is described with a composition that is considered in our study to be a \textit{mortality pattern} (see explanation on page \pageref{mpattern}), it is of primary interest to compare the clusters' patterns with the overall mortality pattern. A comparison of the descriptions of the obtained clusters and the description of the whole data set may be measured with an index comparing each symbolic variable $Y_j$ in each cluster $C$.  Batagelj et al. \cite{Bat-newCh-22} defined a version of such an index, called {\em specificity}. In this work, we change it slightly (by using squared differences instead of absolute values) to agree with the distance used in the clustering process. We define it as
\begin{equation}
	\spec(Y_j,C) = {1 \over 2} \sum_{\ell=1}^{m_j} (r_{\Units j\ell} - r_{Cj\ell})^2, \label{Batagelj:eq:spec}
\end{equation}
where $r_{\Units j\ell}$ is the $\ell$-th component of the symbolic variable $Y_j$ for the leader of the whole set of units $\Units$,  and $r_{Cj\ell}$ is the $\ell$-th component of the symbolic variable $Y_j$ for the leader of cluster $C$. Note that this index shares the same characteristics as the one from \cite{Bat-newCh-22}, namely $0 \leq \spec(Y_j,C) \leq 1$.

To additionally identify the most characteristic components $\ell$ of the symbolic variable $Y_j$ on the cluster $C$, we compute the \emph{contrast} indices (also proposed in \cite{Bat-newCh-22})
\begin{equation} 
\mbox{contrast}_{j\ell} = \begin{cases} \frac{ r_{Cj\ell}}{ r_{\Units j\ell} }   &\mbox{if } r_{Cj\ell}  \geq r_{\Units j\ell} \\
	-\frac{r_{\Units j\ell}}{r_{Cj\ell}}   & \mbox{otherwise} \end{cases} , \quad \ell =1,\ldots,m_j 
\end{equation}	
and select those with the highest absolute values. 
By definition, $|\mbox{contrast}_{j\ell}| \geq 1$. A value of $|\mbox{contrast}_{j\ell}|$ close to 1 suggests that this component has almost the same value/pro\-ba\-bility in cluster $C$ as in the whole set $\Units$ (overall value). The higher the value, the greater the contrast between the values of the cluster leader and the representative of all countries. 

The calculated contrasts and specificities for the three main clusters are given in Table~\ref{cont-spec}, with the contrasts with values at least 1.25 highlighted. We see that the highest contrasts occur for 
suicides, with the second cluster (from left-hand to right-hand in Fig.~\ref{dendro}) having values below the overall value in all age-gender combinations, while in the first cluster all contrasts are above the overall value. We also observe that deaths due to respiratory diseases are more pronounced in the second cluster, especially in the youngest age group, while contrasts for respiratory disease in the other two clusters are below the overall value in all age-gender combinations. Additionally, contrasts for deaths due to circulatory disease are all above the overall value with a value of at least 1.25 in the second cluster; while there seem to be less neoplasms in cluster three. 

Additionally, we explored possible relationships between mortality patterns in the identified clusters and external sociodemographic indicators (which are displayed at the bottom of the dendrogram from Fig.~\ref{dendro}). Comparisons of values between groups with classical one-way ANOVA and post-hoc multiple comparison Bonferroni tests show the most significant differences between the first and the third group (from left-hand to right-hand in Fig.~\ref{dendro}) in the number of \textit{expected} deaths in the entire age group 20-39 years per 100,000 of the standard population 
(CDR $\mbox{mean difference } = -14.04, p < 0.01$). Note that this variable is not completely independent from the clustering variables. A significant difference is also found between the first two clusters in the proportion of government spending in health (GOV $\mbox{mean difference } = 3.34,  p < 0.1$).

\begin{table}[h!]
\caption{Specificities of eight selected age-gender combinations and cause-of-death contrasts for selected three clusters of EU countries}
\begin{center}
{\scriptsize
\begin{tabular}{lcrrrrrrr} \label{cont-spec}
	variable & specificity & Neop & Nerv & Circ & Resp & Acc & Suic & Oth\\ \hline  \vspace*{-0.25 cm} \\
	CLUSTER 1: & & & & & & & & \\ \hline
F.Y20-24 & 0.14 & -1.10 & -1.01 & -1.13 & {\bf -1.46} & -1.04 & {\bf 1.32} & -1.00\\
F.Y25-29 & 0.13 & 1.07 & 1.09 & -1.17 & -1.03 & -1.02 & 1.22 & -1.14\\
F.Y30-34 & 0.21 & 1.11 & 1.04 & {\bf -1.26} & -1.19 & -1.10 & {\bf 1.34} & -1.14\\
F.Y35-39 & 0.16 & 1.08 & -1.02 & -1.17 & -1.21 & -1.02 & {\bf 1.33} & -1.16\\
M.Y20-24 & 0.09 & -1.01 & 1.10 & 1.02 & -1.29 & -1.03 & 1.14 & -1.17\\
M.Y25-29 & 0.17 & 1.12 & 1.16 & {\bf -1.26} & {\bf -1.27} & -1.02 & 1.19 & -1.19\\
M.Y30-34 & 0.26 & {\bf 1.25} & 1.22 & -1.21 & {\bf -1.34} & -1.00 & 1.23 & -1.24\\
M.Y35-39 & 0.30 & 1.24 & 1.21 & -1.18 & {\bf -1.35} & 1.01 & {\bf 1.30} & -1.23\\  \hline  \vspace*{-0.25 cm} \\
	CLUSTER 2: & & & & & & & & \\ \hline
F.Y20-24 & 0.43 & 1.23 & -1.14 & {\bf 1.26} & {\bf 1.76} & 1.00 & {\bf -2.02} & -1.01\\
F.Y25-29 & 0.23 & 1.13 & 1.08 & {\bf 1.27} & {\bf 1.38} & -1.13 & {\bf -1.59} & -1.00\\
F.Y30-34 & 0.29 & 1.11 & 1.15 & {\bf 1.31} & {\bf 1.47} & -1.15 & {\bf -1.79} & -1.11\\
F.Y35-39 & 0.26 & 1.05 & -1.01 & {\bf 1.33} & {\bf 1.54} & -1.04 & {\bf -2.13} & -1.15\\
M.Y20-24 & 0.69 & {\bf 1.38} & 1.13 & {\bf 1.25} & {\bf 2.02} & 1.10 & {\bf -1.75} & -1.06\\
M.Y25-29 & 0.65 & {\bf 1.31} & -1.11 & {\bf 1.36} & {\bf 1.93} & 1.09 & {\bf -1.77} & -1.03\\
M.Y30-34 & 0.37 & 1.19 & -1.10 & {\bf 1.37} & {\bf 1.68} & -1.02 & {\bf -1.57} & -1.06\\
M.Y35-39 & 0.45 & 1.21 & -1.14 & {\bf 1.33} & {\bf 1.57} & -1.06 & {\bf -1.72} & -1.11\\ \hline  \vspace*{-0.25 cm} \\
	CLUSTER 3: & & & & & & & & \\ \hline
F.Y20-24 & 0.04 & -1.03 & 1.11 & -1.01 & -1.07 & 1.06 & -1.15 & 1.01\\
F.Y25-29 & 0.26 & -1.24 & -1.21 & 1.02 & {\bf -1.27} & 1.11 & -1.08 & 1.18\\
F.Y30-34 & 0.47 & {\bf -1.29} & -1.18 & 1.09 & -1.10 & 1.20 & -1.21 & 1.23\\
F.Y35-39 & 0.40 & -1.18 & 1.04 & -1.03 & -1.15 & 1.06 & -1.10 & {\bf 1.28}\\
M.Y20-24 & 0.13 & {\bf -1.31} & {\bf -1.27} & -1.23 & {\bf -1.63} & -1.03 & 1.10 & 1.22\\
M.Y25-29 & 0.12 & {\bf -1.44} & -1.15 & 1.04 & {\bf -1.37} & -1.03 & 1.02 & 1.19\\
M.Y30-34 & 0.22 & {\bf -1.52} & -1.21 & -1.01 & -1.10 & 1.01 & -1.05 & 1.22\\
M.Y35-39 & 0.28 & {\bf -1.51} & -1.15 & -1.04 & -1.06 & 1.02 & -1.06 & 1.23\\
\end{tabular}		
}
\end{center}
\vspace*{-1cm}
\end{table}

\section{Conclusion}

In this paper, we presented alternative clustering methods to identify clusters of EU countries with similar mortality patterns in the young population based on the more informative symbolic data description as one of the possible approaches to analyze mortality data at the aggregate level. In doing so, we considered more comprehensive information on the distribution of deaths among the main causes of death by different age-gender groups. A compositional data approach might also be suitable with this kind of data representation and a comparison between the methods would be reasonable, but both are out of the scope of this paper.

The main advantages of the symbolic description are that this representation retains information at two levels for each age and gender group: the distribution among the main causes of death and also the number of deaths (which can be included as weights of the distributions). The symbolic description of the data requires the analysis tools to be adapted. To this end, we presented the adaptation of compatible leader and agglomerative hierarchical clustering methods.

We applied clustering methods to mortality data of the young EU population and identified the main patterns across causes of death. We presented some results that were obtained with hierarchical clustering with weights. The main differences between the three major clusters in different age-gender combinations were found in different causes of death for suicides, respiratory and circulatory diseases, and neoplasms. 
Some connections with the external sociodemographic indicators were found. To confirm these associations (i.e., mortality patterns and government spending for health), a more detailed quantitative and qualitative analysis should be performed.

\section*{Acknowledgment}

This work was partially supported by the Ministry of Higher Education, Science and Technology of Slovenia, Grants P1-0294 and P3-0154. 


\appendix
\section{Appendix 1}
%

\vspace*{-0.6 cm} 
\begin{table}[h!]
\caption{Symbolic description with eight symbolic variables for the young adults in the whole European Union (causes of deaths that do not fall in any of the first six categories are grouped under the "Oth" category)}	
	\noindent
	\hspace*{-0.2cm} 
	{\scriptsize
	\begin{tabular}{lccccccccc}  
		& \quad  Neop & Nerv & Circ & Resp & Acc & Suic & Oth & Weight \\
		$\y\textsubscript{EU28, 20-24 male} $ & $ = [$ 0.093,&0.046,&0.054,&0.022,&0.387,&0.218,&0.180 $]$	 & 2.305\\	
		$\y\textsubscript{EU28, 25-29 male} $ & $= [$ 0.104,&0.037,&0.079,&0.023,&0.322,&0.214,&0.221 $]$ & 2.734\\   
		$\y\textsubscript{EU28, 30-34 male} $ & $ = [$ 0.127,&0.033,&0.118,&0.026,&0.255,&0.182,&0.259 $]$	 & 3.324\\	
		$\y\textsubscript{EU28, 35-39 male} $ & $= [$ 0.154,&0.030,&0.160,&0.033,&0.200,&0.145,&0.278 $]$ & 4.358\\   
		& & & & & & & & \\
		$\y\textsubscript{EU28, 20-24 female} $ & $= [$ 0.168,&0.066,&0.082,&0.040,&0.245,&0.146,&0.252 $]$ & 0.757\\   
		$\y\textsubscript{EU28, 25-29 female} $ & $= [$ 0.241,&0.055,&0.090,&0.038,&0.163,&0.128,&0.284 $]$ & 0.961\\   
		$\y\textsubscript{EU28, 30-34 female} $ & $= [$ 0.328,&0.043,&0.110,&0.036,&0.123,&0.102,&0.259 $]$ & 1.310\\   
		$\y\textsubscript{EU28, 35-39 female} $ & $= [$ 0.411,&0.038,&0.122,&0.033,&0.087,&0.074,&0.235 $]$ & 1.956 
		\\   
	\end{tabular}
	}
\end{table}

\vspace*{-0.8 cm}
\section{Appendix 2}

For the derivation of the leader (i.e., the optimal cluster's representative to the constrained problem), we assume that the symbolic variables are independent so that we can optimize separately for each symbolic variable $Y_j, j = 1, ..., p,$ and therefore omit the index $j$ from the derivations.
We want to find such a vector with non-negative components $\leadl = [ \rr_{1},\ldots,\rr_{m} ]$ with the constraint $\sum_{\ell=1}^{m} \rr_{\ell} = 1$, which minimizes the following function 
$ F(\leadl) = \sum_{X \in C} w_{x} \sum_{\ell=1}^{m} (x_{\ell} - \rr_{\ell})^2$, 
where $\sum_{\ell=1}^{m} x_\ell = 1 \mbox{ for all units } X$.
Using the Lagrange multiplier method, we have to solve the following optimization problem:
$ 
\min_{\leadl}  \left[ F(\leadl) - \lambda \sum_{\ell=1}^{m} \rr_{\ell} \right]. $
With the partial derivation to $\rr_{\ell}$, we get  
\begin{equation*}
\sum_{X \in C} w_{x} 2 (x_{\ell} - \rr_{\ell}) (-1) - \lambda = 0 \mbox{ i.e.,  } 
\rr_{\ell} = \frac{\sum_{X \in C} w_{x} x_{\ell}}{\sum_{X \in C} w_{x}} + \frac{\lambda}{2 \sum_{X \in C} w_{x}}
\end{equation*}
for each $\ell, \ell = 1, ..., m$.
Considering the constraint $\sum_{\ell=1}^{m} \rr_{\ell} = 1$, we get $\lambda = 0$ and thus we obtain for the $\ell$-th component of the optimal representation of the cluster $C$ with 
$\displaystyle \rr_\ell = \frac{1}{\sum_{X \in C} w_{x}} \sum_{X \in C} w_x x_\ell.  $

\end{document}